# Exploring the Public Reaction to COVID-19 News on Social Media in Portugal

Luciana Oliveira[1], Arminda Sequeira[1], Adriana Oliveira[1], Paulino Silva[1] and Anabela Mesquita[2]
[1]CEOS.PP ISCAP Polytechnic of Porto, S. Mamede de Infesta, Portugal
[2] ISCAP Politécnico of Porto & Algoritmi RC, S. Mamede de Infesta, Portugal
lgo@eu.ipp.pt
arminda@iscap.ipp.pt
aoliveira@iscap.ipp.pt
paulino@iscap.ipp.pt
sarmento@iscap.ipp.pt

**Abstract:** The outburst and proliferation of the COVID-19 pandemic, together with the subsequent social distancing measures, have raised massive challenges in almost all domains of public and private life around the globe. The stay-at-home movement has pushed the news audiences into social networks, which, in turn, has become the most prolific field for receiving and sharing news updates, as well as for public expression of opinions, concerns and feelings about the pandemic. Public opinion is a critical aspect in analysing how the information and events impact people's lives, and research has shown that social media data may be promising in understanding how people respond to health risks and social crisis, which are the feelings they tend to share and how they are adapting to unforeseen circumstances that threaten almost all societal spheres. This paper presents results from a social media analysis of 61532 news headlines posted by the major daily news outlet in Portugal, Sic Notícias, on Facebook, from January to December 2020, focusing on the issues attention cycle and audiences' emotional response to the COVID news outburst. This work adds to the emergent body of studies examining public response to the coronavirus pandemic on social media data.

## 1. Introduction

Covid-19 has been declared a global pandemic on the March 11th, 2020, by the World Health Organization (WHO). By the end of May 2020, the virus had spread all over the globe, with more than 4 million confirmed cases of infection leading the WHO to declare a Global Health Emergency. As of early 2021, the situation still develops, and although the vaccination process has already begun, at least in the developed countries, there is no foreseeable end to the pandemic. Everyday activities around the world were suspended or restricted, and people were confined at home in an unprecedented situation, completely unprepared and without knowing how the scenario will evolve.

The stay-at-home movement pushed outbreaks of news into social media, as audiences had easy access to information otherwise limited, considering the traditional channels. In the present context, the preamble to the constitution of World Health Organization gains relevance particularly when it highlights the importance of accurate information to citizens about health issues: "informed opinion and active cooperation on the part of the public are of the utmost importance in the improvement of health of the people"(WHO, 2002). In this context and for that reason, social media plays an important role in bridging people and news providers, not only by permitting the massive flow of health, normative, political, and social information but also by providing the means for audiences' expression of opinions and emotions. In fact, social media became a prolific field for receiving and sharing news updates and for public expression of opinions, concerns, and feelings about the pandemic. Social media has become a central platform for communication and interaction, as well as a lens to the current social reality. As such, it is an important container of information that has great value both for businesses and for social purposes (Sharma & Dey, 2012).

The public expression of audiences' opinions and emotions on Social Media allows to estimate the impacts of events and news on people's lives, by understanding how audiences relate to the subject, and how audiences' emotional state has fluctuated. Collective emotions result from a large number of individuals sharing one or more emotional states, which can emerge in online communities (Kappas, 2017), and foster emotional contagion (Ferrara & Yang, 2015). As collective emotions tend to sustain themselves for longer periods than individual emotional reactions (Garcia, Kappas, Küster, & Schweitzer, 2016), knowledge about the general population can assist in detecting abnormal affective dynamics, which have been linked to mental disorders like depression (Koval, Pe, Meers, & Kuppens, 2013).

Alongside, the type and volume of news and of audiences' interactions with them allow to analyse the amount of attention that an issue attracts, both by the mass media and among audiences, determining the issues attention cycles (Downs, 1972) throughout time, as the situation seems to perpetuate. As stated by Giuntini et al. (2019), the large sharing of data in social networks has fostered the analyses of human behaviour facilitated by the fact that social networks have become an environment in which users feel comfortable sharing their particularities such as their ideas, thoughts, and opinions. In fact, social media consists of a fruitful environment for the investigation of behaviour, social interactions (Cioban & Vîntoiu, 2020), and how people adapt to unforeseen circumstances and threats, and the texts, images, videos, reactions, and other forms of interaction that users share are able to provide an added lens into, for instance, mental health issues. Additionally, Rafi, Rana, Kaur, Wu, and Zadeh (2020) state that analysing public opinion on social media may be critical in determining a sense of social well-being and public response to the crisis as a first step to solving emergent problems.

The main goals of this work consist of analysing how COVID-19 news were framed in the journalistic activity of 2020, how did audiences engage with news overall and with COVID-19 news specifically, and which was the audience's emotional response along the year. We perform a temporal analysis and study variations in the attention devoted to COVID-19 news and in the audiences' emotional response during 2020, from before the first case of infection was reported (pre-pandemic) to the moment when the Portuguese government announced the first vaccination measures, in late December.

## 2. Background

Research devoted to audiences' response to COVID-19 on social media has gained momentum since the early stages of the global pandemic. In this section, we identify some of the relevant work associated with categorical emotion detection, issues attention cycle and framing.

On sentiment analysis Burzyńska, Bartosiewicz, and Rękas (2020), performed a research in Poland, about the frequency and impact of online mentions about the COVID-19 illness taken from social media platforms (Facebook, Instagram, Twitter, blogs, forums) to highlight and better understand the scope of coronavirus discussion in the country. They used SentiOne social listening tool to gather the data and perform the monitoring between February 24$^{th}$ 2020 to March 25$^{th}$ 2020. Cioban and Vîntoiu (2020), in Romania, developed an investigation on headlines with text bodies and comments posted on Reddit, using word frequency and sentiment, analysing feelings expression towards the context of social crisis during the most restrictive times during the pandemic.  Zou, Wang, Xie, and Li (2020) analysed the public reactions (English language) on Twitter about the COVID-19 pandemic comparing the United States and United Kingdom from March 6$^{th}$, 2020 to April, 2$^{nd}$ 2020. The sentiment scores of the tweets on COVID-19 were analysed and associated with the policy announcements and the confirmation of cases in both countries. X. Wang, Zou, Xie, and Li (2020), on the other hand, analysed the changes and key themes discussion topics and sentiments (English language), on Twitter, in California and New York (United States), collected from March 5$^{th}$ 2020 to April 2$^{nd}$ 2020.  Chandrasekaran, Mehta, Valkunde, and Moustakas (2020) examined key terms and topics Covid-19-related and sentiment changes over time from January 1$^{st}$ to May 9$^{th}$, 2020 to uncover key trends. T. Wang, Lu, Chow, and Zhu (2020) performed research on Sina Weibo, a popular Chinese social media, analysing posts with negative sentiment to assess public concerns. 999,978 randomly selected COVID-19 related Weibo posts from January 1$^{st}$ 2020 to February 18$^{th}$ 2020 were analysed. Li, Xu, Cuomo, Purushothaman, and Mackey (2020) conducted a quantitative and qualitative assessment of Chinese social media posts originating in Wuhan City on the Chinese microblogging platform Weibo during the early stages of the COVID-19 outbreak. On audiences' engagement Marivate, Moodley, and Saba (2020) analysed how the law and regulation promulgated by the government in response to the pandemic contrasts with discussion topics social media users have been engaging in, based on twitter posts, during the period of March 1$^{st}$ to May 17$^{th}$ 2020. Ofoghi, Mann, and Verspoor (2016) explored emotion classification of Twitter microblogs related to localised public health threats and studied whether the public mood can be effectively utilised in early discovery or alarming of such events.

### 2.1 Issue-attention cycle

The concept of issue-attention cycle refers to the ups and downs of attention an issue receives either from the public or from mass media (Downs, 1972) taking into consideration that, for most of the issues, media attention and audience's attention don't hold for a long period of time. The "cycle" proposed by Downs comprises five phases (Figure 1). First is the pre-problem phase in which an issue does not capture much public attention of the audiences. At this stage only a fraction of people, such as experts or interest groups, are aware of it. In the second phase, public awareness raises, and a period of alarmed discovery associated with specific risks may occur. However, this is often accompanied by the optimistic belief that, by taking some measures, the problem will be solved. When people begin to realise that solving the problem is beyond their initial assessment and maybe wider and much more resources consuming, we reach the third stage. The fourth phase is characterised by the gradual decline in public interest in the problem and a certain disconnection, although the problem may persist. In this final phase, an issue is replaced by other concerns and is subject to "spasmodic recurrences of interest" (Downs, 1972, pp. 39-40).

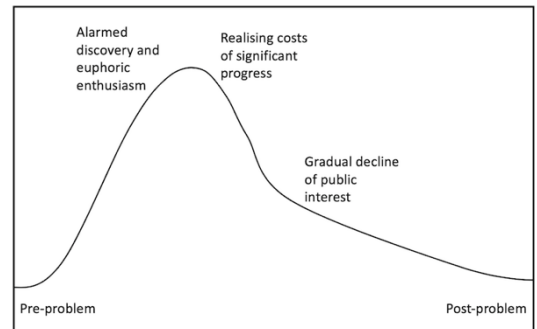

**Figure 1:** Issue Attention Curve (Rose, 2020)

Although there is some criticism concerning the linear nature of Down's model, especially because it doesn't take into consideration the dynamics of social interactions as well as social and cultural factors contributing to the problem's rise and fall in attention, media tend to construct problems linearly framing issues as having beginnings, middles and ends in a generally one-directional temporal fashion (McComas & Shanahan, 1999). The authors propose another issue attention cycle model, embedded in the narrative communication theory[1], stating that narratives use specific temporal order of events to construct meanings, being their model divided into three phases: the waxing phase, the maintenance phase, and the waning phase (McComas & Shanahan, 1999). Each phase shows differences in narratives salience. In the waxing phase, the media attention increases and the focus on consequences and implied danger are more common. In the maintenance phase, attention remained relatively constant and, the frame focuses more on controversy among scientists. Finally, in the waning phase, consequences are highlighted but the focus on the theme progressively fades away from the news.

## 2.2 Emotional responses on social media

It is primarily important to differentiate emotion detection from sentiment analysis. Sharma and Dey (2012) state that sentiment analysis is a process of studying public opinion about an issue. Balahur (2013) defines sentiment analysis as the Natural Language Processing (NLP) task of dealing with the detection and classification of sentiments in texts. It consists of a process of classification of texts in classes considered "positive", "negative" and "neutral", and it can be fine-grained into subcategories such as "very positive" and "very negative", when considering its representation in values ranging from -1 to 1. As seen in the previous sections, this technique is very popular in analysing the public response to COVID-19 on social media.

Emotion detection comprises the task of classifying text into several classes of emotion. Some of the research in the field has found sentiment analysis and emotion detection under the umbrella of sentiment analysis, albeit distinct in some respects, (Balahur, 2013). Emotions, however, are much more expressive than sentiments, as they do not need to contain a sentiment and vice-versa (Liu, 2012; Y. Wang & Pal, 2015). On Facebook, users often adopt the use of emoticons in posts, messages, and comments to increase the meaning of these messages and express emotions with symbols without the need to write. Emoticons are small images or combinations of diacritical symbols, intentionally developed to replace nonverbal components of communication, suggestive of facial expressions (Giuntini et al., 2019), and consist of a categorical emotion model. On social media, emoticons have become the most widely adopted means of expressing emotions (Oleszkiewicz et al., 2017). There are several studies devoted to analysing the sentiment of emojis and Facebook reactions (Cazzolato et al., 2019; Giuntini et al., 2019; Tian, Galery, Dulcinati, Molimpakis, & Sun, 2017). The Facebook reactions, an expansion of the "Like" button, were introduced on February 24th, 2016, and in 2020 during the COVID-19 outbreak, the "Care" reaction was added to the set (Figure 2).

---

[1] Narrative theory generally argues that humans use narratives to weave together fragmented observations and perceptions to build meanings and realities.

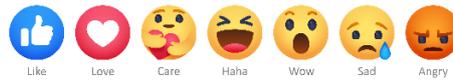

**Figure 2:** Facebook reaction set, as of early 2020

Emotion models are essentially of two types: dimensional and categorical (Y. Wang & Pal, 2015). The dimensional models represent emotions on three temporal dimensions: valence, arousal and dominance (Ekkekakis, 2013). A recent example is the model of affective dynamics on social media, proposed by Pellert, Schweighofer, and Garcia (2020). Categorical emotion models assign emotions to predetermined categories. The most popular emotion taxonomy is Ekman's basic emotion set: anger, disgust, fear, happiness, sadness, and surprise (Ekman, 1992). The author views emotions as discrete, automatic responses to universally shared, culture-specific and individual-specific events (Ekman & Cordaro, 2011). Ekman's work has been used in several works using automatic classification of social media textual data to understand public mood. For instance, Ofoghi et al. (2016) has focused on emotions related to Ebola incidents on Twitter, and X. Li et al. (2020) explored cultural emotional differences between America and China to depict the dynamics of public affection in the time of COVID-19. Giuntini et al. (2019) state that the use of emoticons as universal emotions should enable methods of emotional analysis to be employed with minimum dependence on verbal semantics. This is of particular relevance to our work because we use Facebook reactions as a core function to analyse audiences' emotional response to COVID-19 news posted on Facebook in 2020. The authors analyse the valence of Facebook reactions and their ability to describe Ekman's basic emotions and conclude that the attribution of emotions and polarity indicates that there might be a relationship between the emotions felt and the reactions expressed in the virtual environment. Findings reveal that the reactions "Angry" and "Sad" have negative polarity, "Love" has positive valence and "Haha", "Wow" and "Like" are volatile, acquiring meaning together with other reactions. Regarding basic emotions, there are some strong associations with reactions. "Angry" expresses anger, "Wow" expresses surprise, "Sad" represents sadness and "Love" represents joy. "Haha" appears to express some degree of surprise and joy and "Like" is ambiguous in terms of polarity and emotion. The only basic emotion with no observable representation in reactions is fear (Giuntini et al., 2019).

## 3. Methods and procedures

This work follows the general approach of quantitative content analysis (Bryman, 2016), and it consists of a descriptive study. We used the Facebook Graph API to download the news posted by the major daily news provider in Portugal, Sic Notícias, between the January 1st and December 31st 2020. This page has more than 1,86 million followers on Facebook (more than 18% of the Portuguese population). The dataset is composed of 61532 news posted on the network, for which we collected the *created date* and *time*, *link* (news external URL), *message* (text included in the post), *link text* (the title of the news), *description* (news lead), *likes*, *comments*, *shares*, *love*, *wow*, *haha*, *sad*, *angry* and *care*. We refer to "Like", "Comment" and "Share" as forms of interaction with content; and to "Love", "Wow", "Haha", "Sad", "Angry" and "Care" as reactions, in the sense that these convey emotional responses. Our assumption is that the "Like" button is somewhat a default type of interaction with content. Similarly, Freeman, Alhoori, and Shahzad (2020) state that the expansion of the "Like" button requires more effort to register a special reaction and that, since 2016, Facebook users were well accustomed to using "Likes" to respond to a variety of content, namely positive and negative posts, which was also observed by Giuntini et al. (2019). In fact, in our dataset the average of "Likes" is, in most cases, one order of magnitude higher than most special reactions and two times higher than other forms of interaction with content, as shown in Table 1; thus it is not used as a variable for recording the emotional response.

**Table 1:** Descriptive statistics of interactions and reactions

|      | Likes  | Comments | Shares | Love  | Wow   | Haha  | Sad    | Angry | Care  |
|------|--------|----------|--------|-------|-------|-------|--------|-------|-------|
| Min. | 0      | 0        | 0      | 0     | 0     | 0     | 0      | 0     | 0     |
| Max. | 28,080 | 20,712   | 44,428 | 5,469 | 2,397 | 1,647 | 9,456  | 5,886 | 1,297 |
| Mean | 110.91 | 51.10    | 45.65  | 8.02  | 7.44  | 11.79 | 27.27  | 11.65 | 1.54  |
| σ    | 452.88 | 158.77   | 348.10 | 72.30 | 37.11 | 48.93 | 133.66 | 80.20 | 13.62 |

The dataset of news was categorised into two subsets: COVID-19 related news and Other news. We decomposed the news external URL path, for the posts type "Link", according to the journalistic themes used by the news provider. Example for COVID-19 news: https://sicnoticias.pt/*especiais*/*coronavirus*/Casos-confirm[...]. Example for the category Other news: https://sicnoticias.pt/*economia*/Projeto-piloto-para[...]. The remaining post types (Photo, Live Video Complete, Native Video and Status) were manually categorised by the authors. Descriptive statistics of post types for the subsets of news are presented in Table 2.

**Table 2:** Descriptive statistics of post type per subset (category of news)

|  |  | COVID-19 | | Other | |
| --- | --- | --- | --- | --- | --- |
|  |  | Count | N % | Count | N % |
| Post type | Link | 26,319 | 99.5% | 34,436 | 98.2% |
|  | Photo | 45 | 0.2% | 183 | 0.5% |
|  | Live Video Complete | 8 | 0.0% | 18 | 0.1% |
|  | Native Video | 75 | 0.3% | 443 | 1.3% |
|  | Status | 2 | 0.0% | 3 | 0.0% |
| Total news (Count, %) |  | 26,449 | 42.98% | 35,083 | 57.02% |

An additional dataset containing the confirmed cases of COVID-19 infection in Portugal, in 2020, was obtained from the WHO website[2] and was used to set the context for the analysis of news and audience response.

## 4. Results

In this section, we present the results regarding the framing of COVID-19 news in the journalistic activity and subsequent audiences' response to COVID-19 news.

### 4.1 Framing COVID-19 news in the journalistic activity

As seen in Table 2, COVID-19 news represent nearly 43% of all news posted on Facebook in 2020. The distribution of the two categories of news per week/month is presented in Figure 1, together with the national context for confirmed COVID-19 infection cases and key moments.

Figure 3 shows the two COVID-19 waves in Portugal, preceded by the pre-COVID-19 stage up until week 9. The first wave between weeks 10 to 35, and the second between weeks 36 and 53. There is a statistically significant variation in the volume of COVID-19 news over the months of the year ($\chi2$ (11)=10900.113; p<0.01). January (5.9%) and February (16.6%) are the months with the lowest volume of COVID-19 news, and March (73.6%), April (77.6%) and May (60.8%) are the months in which the highest relative percentage is recorded, particularly between weeks 11 and 21. This consists of the beginning of the first wave, with the outburst of infections, first confinement measures, first State of Emergency, total lockdown, first Stage of Contingency and first phases of reopen.

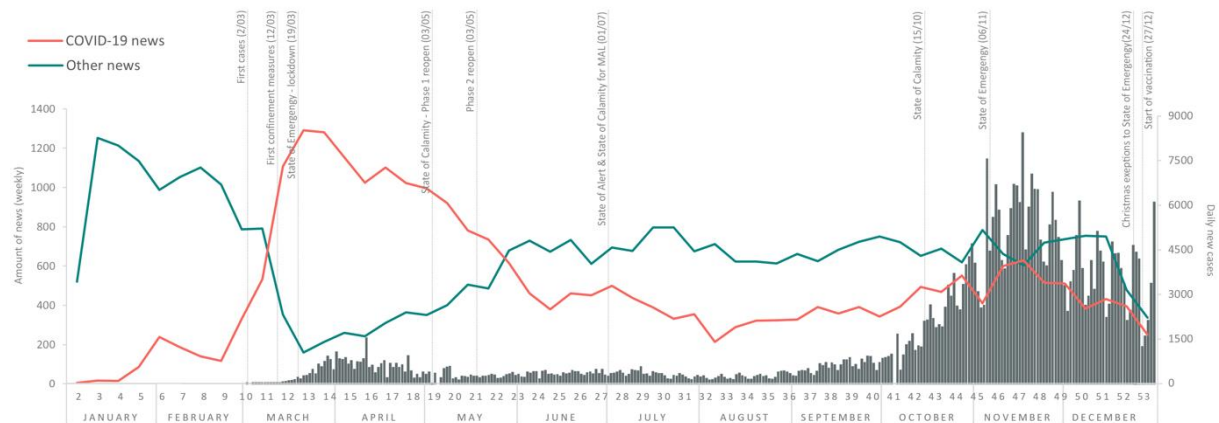

**Figure 3:** Overview of news, confirmed COVID-19 cases and key moments

---

[2] WHO Coronavirus Disease (COVID-19) Dashboard, available in https://covid19.who.int/

In the remaining months of the year, the volume of COVID-19 news fluctuated between 30.8% (August) and 42.6% (November). The lowest peak of COVID-19 news after the lockdown, and a bigger gap between the two categories of news, was registered in week 32, after a decline of five weeks in a row since week 28. This period precedes the month with the lowest number of daily infections (weeks 32 to 35), which also consists of the holiday season for most workers and the State of Alert. From here onwards, despite the drastic increase in the cases of infection (second wave), the volume of COVID-19 news and audiences' emotional expression suffers an overall minor increase, with similar fluctuations, never reaching the same volumes observed in the first wave. There is a single moment in week 47 when the volume of COVID-19 news (51,01%) surpasses the volume of Other news. The variation of COVID-19 news along the year is not, however, substantially correlated with the reported daily cases of infection, and this is particularly visible in the last trimester. In fact, the evolution of daily cases only explains 3,4% of the variation of COVID-19 news in 2020 (F (1, 50715)=1804,669; p<0,001).

Regarding the days of the week, although there are no significant variances in the average of post per day of the week among the two subsets of news, ($t$ (61530)=0,406; p>0,05), it was possible to observe that the day of the week in which more news are posted, overall, is Wednesday, and that Sundays are the days in which the least overall amount of news are posted on social media (Table 3). Considering the entire period (one year), the total percentage of the category "Other" news is more frequent that COVID-19 related news, with prevalence of "Other news" on Wednesdays (59,2%) and of COVID news on Saturdays (44,4%) ($\chi^2$ (6)=34,806; p<0,01). When considering only the first 11 weeks (weeks 11-21) since the first confinement measures, there is a reversed tendency. Posts related to COVID-19 vary between 73,1% (min.) and 79,9% (max.) in all days, whereas other news varies between 20,1% (min.) and 26,9% (max.). Saturday is also the day registering the highest percentage of COVID-19 news (79,9%) and Wednesdays the lowest percentage (73,1%) ($\chi^2$ (6)=37,096; p<0,01).

**Table 3:** Covid and Other news per day of the week

|  | %Total news | % News (all weeks) |  | % News (weeks 11-21) |  |
|---|---|---|---|---|---|
|  |  | COVID-19 | Other | COVID-19 | Other |
| Sunday | 8,54 | 43,9 | 56,1 | 77,3 | 22,7 |
| Monday | 16,20 | 44,0 | 56,0 | 74,0 | 26,0 |
| Tuesday | 16,41 | 42,3 | 57,7 | 74,6 | 25,4 |
| Wednesday | 17,20 | 40,8 | 59,2 | 73,1 | 26,9 |
| Thursday | 16,51 | 42,9 | 57,1 | 77,8 | 22,2 |
| Friday | 15,67 | 43,8 | 56,2 | 76,3 | 23,7 |
| Saturday | 9,47 | 44,4 | 55,6 | 79,9 | 20,1 |

### 4.2 Audiences' response to COVID-19 news

In this section, we analyse audiences' engagement (Figure 3) with news regarding intensity (volume) and emotional response, consisting of click-based special reactions (Figure 4).

The variation of audiences' engagement throughout the year is explained by the presence of COVID-19 news, since the average of all interaction types is lower for Other news ["Like" ($t$(61277,657)=13,709; $p$<0,001; "Comments" ($t$(40887,612)=6,577; $p$<0,001; "Shares" ($t$(33162,910)=11,832; $p$<0,001]. The most frequent interaction type every month is "Like" followed by "Share" and "Comment" ["Like" ($H$ (11)=1771,254; p<0,001); "Share" ($H$ (11)=1111,402; p<0,001); "Comment" ($H$ (11)=450,337; p<0,001)], except when considering March, when "Share" exponentially peaked in week 11 regarding COVID-19 news ($t$(1362,633)=4,284; $p$<0,005). March, April, and May are the months in which audiences have mostly interacted with news content. March and April is also when audiences have expressed a higher volume of emotional responses (*SumReactions*), particularly in weeks 11 to 13, leading us to believe that it is in response to the first confinement measures and subsequent lockdown (1st State of Emergency) ($H$ (11)=303,246; p<0,001). August, September, December and February are the months in which audiences have least engaged with content, considering all interaction types ($H$ (11)=869,571; p<0,001). During these months there is an increase of Other news in detriment of COVID-19 news, as the country is in pre-pandemic, moves to deconfinement or is on holidays period. From August onwards, as the ratio of COVID-19 news begins to increase again, the audiences' overall engagement also increases, mostly through "Likes" and "Comments". Between October and December, there is also a rise in "Reactions" and "Shares" when compared to the four previous months.

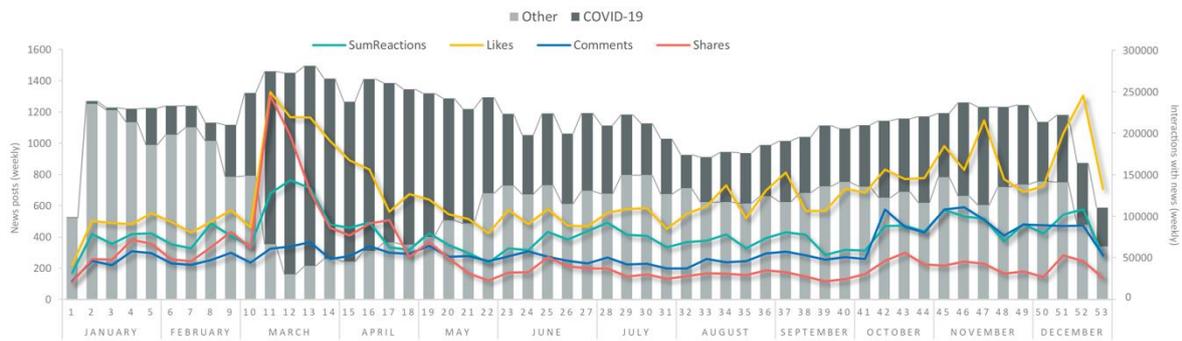

**Figure 4:** Evolution of audiences' interaction with content

The audiences' emotional response (click-based special reactions) is depicted in Figure 4, evidencing a noticeable and persistent feeling of sadness ("Sad"), that is greatly explained by the presence of COVID-19 news ($t(34942,047)12,496$; $p<0,001$).

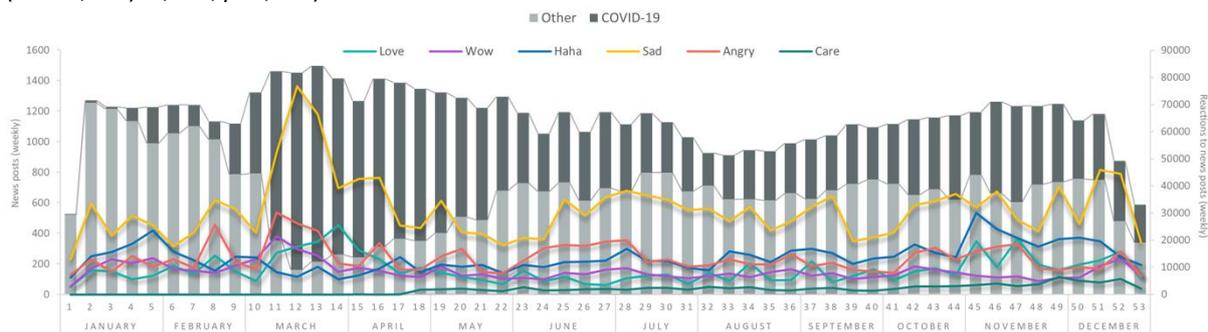

**Figure 5:** Evolution of audiences' emotional response

The audiences' emotional response is also explained by the presence of COVID-19 news for the emotions "Wow" ($t(35141,003)=9,442$; $p<0,001$) and "Care" ($t(14225,920)=5,948$; $p<0,001$], but not for the emotion "Love" ($t(28250)=-1,370$; $p=n.s.$) and "Angry" ($t(29839)=-0,824$; $p=n.s.$). Although without statistical significance, these latter emotions are expressed on average more towards Other news. The emotions "Wow", "Sad", and "Care" tend to increase together with the volume of COVID-19 news, while the emotion "Haha" tends to decrease ($t(35481,246)=-8,112$; $p<0,001$). The overall average of "Sad" reactions to COVID-19 news (59,22) is one and a half times higher than for Other news (17,55) and is much higher than any other emotional reaction to both categories of news.

There is no correlation between the evolution of infection cases and the expression of sadness ($r=0,01$), but there is significant variation on the average of "Sad" reactions per week ($H(52)=459,363$; $p<0,001$), with two peaks in weeks 12 (52,92) and 52 (51,02), which consist of the two weeks preceding the declarations of State of Emergency. The saddest periods of the year are, in decreasing order, from weeks 11 to 26 (first weeks of the lockdown), weeks 42 to 53 (since the declaration of the State of Calamity to the end of the year) and from weeks 25 to 38. We specifically analysed the presence of sadness in the pre-pandemic period (weeks 1-10) to evaluate the possibility of it being a result of negative events reported in Other news, which is frequent (Soroka, 2006; Soroka, Young, & Balmas, 2015). Results show that, in this period, the average of "Sad" reactions to COVID-19 news (47,22), when compared to Other news (41,63), is not significant, thus sadness in this period must not be attributed specifically to COVID-19 news ($t(5898)=0,936$; $p=n.s.$), contrarywise to the remaining weeks of the year.

Regarding expressions of anger, there are two peaks of "Angry" reactions during the year. The second and highest, between weeks 11 to 13, corresponds to the first three weeks of the first confinement and lockdown. The first anger peak, in week 8, consists of an exception. Anger is primarily expressed in Other news, specifically in a subset of five news (50,7% of all "Angry" reactions) related to children and animal deaths due to malnutrition and bad living conditions. The COVID-19 news registering higher "Angry" reactions in this week are related to the Diamond Princess cruise ship, where one of the first massive outbreaks of the new coronavirus took place.

In week 5 the peak is registered for the emotion "Haha" tends towards Other news, but with no statistical significance ($t$ (665)=-1,461; $p$=n.s.), whereas in week 45 it is definitely mostly expressed towards Other news ($t$ (785)=-4,267; $p<0,001$). There is also a peak of "Love" reactions in week 14, with a higher average in COVID news, but no statistical significance ($t$ (672)=0,943; $p$=n.s.). Since audiences tend not to express "Love" in COVID-19 news, we believe that this is consistent with a set of forty positive news about recovered cases, donation and public purchase of health equipment and a reported decrease in the number of infections during this week.

### 4.3 Mass media and audiences' cycles of attention

Having analysed the fluctuations in the volume of COVID-19 and Other news as well as the audiences' emotional responses, we revisit the issues-attention cycle. Analysing Figures 3 and 5 it is possible to observe that there is a great symmetry between the fluctuation of the volume of COVID-19 news and the audiences' emotional response with Downs' the issue-attention cycle (Figure 1). However, it is possible to identify small discrepancies between the attention of mass media to COVID-19 news and the emotional response shared by the audiences. The pre-problem stage, in the national context, for the mass media, lasts until week 8 with lower attention to the pandemic on news and lower reactions.

The second stage – public awareness, alarmed discovery, and euphoric behaviour – is visible between weeks 9 and 12 for the mass media, when immediate measures were applied and communicated to mitigate the spread of the virus. Stage two for the emotional response from audiences begins with delay, about two weeks later (week 11), lasting about one week (until week 12). This is a stage of abundance of news/information and increase of most reaction, mainly sadness, anger, and surprise, but also of heavy information sharing among social media users. In fact, aroused emotional states tend to lead to information sharing in social media (Berger, 2011).

The third stage – realising the cost of significant progress – is noticeable between week 13 and 16 for the mass media, when there is still a high volume of COVID-19 related news, and the end of the lockdown approaches. For the emotional audience's response, this stage is much shorter (weeks 12-13) since there is a decrease in the critical emotions and interactions (excluding "Likes"). As stated by Garcia et al. (2016), emotions relax over time if they are not re-stimulate. We hypothesise that the COVID-19 critical information was shared by the mass media in Stage two and beginning of Stage three, slowly decaying thereafter in Stage 4 (weeks 14-22).

We believe that mass media attention to the problem reaches Stage 5 - post-problem – in week 22, and the audiences' emotional response reaches this stage in week 17, although the actual problem persists, and number of COVID-19 infections greatly increases towards the end of the year. As stated by Downs (1972), in this stage, the issue has been replaced at the centre of public concerns and moves into a prolonged limbo of spasmodic recurrences of interest. After week 22, this noticeable for the mass media attention, when the ratio of COVID-19 news is no longer higher than Other news. For the audiences' emotional response this stage begins sooner, in week 17 considering most reactions. One of the most visible recurrences of interest occurs in week 47, when the volume of COVID-19 news (51,01%) surpasses the volume of Other news again, although there are other less expressive moments until the end of the year. It is also worth noticing that in October, November and December, there is an increase in comments to approximately the same levels observed between April and July. Overall, audiences' show shorter time spans of attention both in terms of emotional response and sharing information on social media.

### 5. Conclusion

We used well-established models and emergent practices for presenting a temporal analysis and studied the variations in the media attention and collective emotions in response to COVID-19 news posted on Facebook, during 2020.

Regarding the Portuguese context, it is possible to observe, first and foremost, an unequivocally sad ("Sad") nation throughout 2020, aggravated by the COVID-19 pandemic. Audiences also show surprise ("Wow") and a supportive, empathic, and companionate attitude ("Care") regarding the health crisis. In Portugal, concerns related to the preservation of mental health have gained media attention, and our work contributes as evidence to support these concerns. In February 2021, the doctor in charge of intensive care unit of Hospital de Santo António stated that there are at least four concurrent crises in Portugal: the COVID19, the non-COVID-19, the economy and the mental health crisis, being that the latter may suffer the repercussions of the first three silently

(Marques da Silva, 2021). In January 2021, a draft resolution had already been presented in the Portuguese Parliament recommending the government to "promote a greater involvement of primary health care in the prevention and treatment of depressive and mood disorders, through consultation, in each of the health centres, specifically dedicated to early diagnosis" (Lusa, 2021).

The fluctuation of mass media attention to COVID-19 news is not particularly determined by the daily evolution of infection cases, leading us to believe that the political, economic, and social repercussions of the health crisis play a major role in the volume of news shared on social media, particularly considering that there is usually only one or two daily pieces of news reporting the evolution of infections. In fact, the peak of COVID-19 news registered in March-April (1st wave) was not observed in subsequent stages, particularly in the last trimester of 2020 (2nd wave) when the cases of infection have more than tripled. This is in line with Downs (1972) issues-attention cycle theory, namely the Phase 2, of alarming discovery. We also believe that this is related to the briefing effect of the first outburst of COVID-19 news, during which there is a higher volume of news and relevant information to communicate about unknown multilevel circumstances, on the mass media side, matched with a higher demand for critical information for decision-making on the audiences' side (individuals and organisations).

We have also identified that, in general, audiences show shorter cycles of attention and emotional reaction to COVID-19 news, based on the total amount of interactions with this content. Although this was expected and identified in previous literature (e.g., in Garcia et al. (2016)), we highlight that it may compromise the efficacy of preventive health measures in the context of a health crisis. We believe that the current emotional exhaustion generated around the COVID-19 infodemic may particularly aggravate the reduction of the audiences' attention cycles after a certain point. Consequently, the audiences' urges to avoid constantly receiving bad or depressing news (information avoidance) may compete with the need to receive relevant information and undermine the recurring appeals to avoid risky health behaviours (compliance avoidance).

This work is not without limitations. We do not present much detail regardings specific moments because our purpose was to provide an overall picture of the entire year, focusing on attention cycles an emotional response. We are aware of the particular interest that an in-depth analysis of the content of news in specific or critical moments could add tremendous value in clarifying the motivation behind the audiences' emotions. As future research, we intend to extend this analysis to the first three months of 2021 as it corresponds to the third wave and most shocking period in the number of infected people and deaths in Portugal.